\documentclass[twocolumn,showpacs,preprintnumbers,amsmath,amssymb,aps]{revtex4}

\usepackage{graphicx}
\usepackage{dcolumn}
\usepackage{bm}

\begin{document}


\title{Magnetoresistance oscillations in multilayer systems - triple quantum wells}

\author{S. Wiedmann,$^{1,2}$ N. C. Mamani,$^3$ G. M. Gusev,$^3$ O. E. Raichev,$^4$ A. K. Bakarov,$^5$
and J. C. Portal$^{1,2,6}$} 
 \affiliation{$^1$LNCMI-CNRS, UPR 3228, BP
166, 38042 Grenoble Cedex 9, France} \affiliation{$^2$INSA Toulouse,
31077 Toulouse Cedex 4, France} \affiliation{$^3$Instituto de
F\'{\i}sica da Universidade de S\~ao Paulo, CP 66318 CEP 05315-970,
S\~ao Paulo, SP, Brazil}\affiliation{$^4$Institute of
Semiconductor Physics, NAS of Ukraine, Prospekt Nauki 41, 03028,
Kiev, Ukraine} \affiliation{$^5$Institute of Semiconductor Physics,
Novosibirsk 630090, Russia}\affiliation{$^6$Institut Universitaire
de France, 75005 Paris, France}
\date{\today}

\begin{abstract}
Magnetoresistance of two-dimensional electron systems with several occupied 
subbands oscillates owing to periodic modulation of the probability of 
intersubband transitions by the quantizing magnetic field. In addition to 
previous investigations of these magneto-intersubband (MIS) oscillations in 
two-subband systems, we report on both experimental and theoretical studies 
of such a phenomenon in three-subband systems realized in triple quantum wells. 
We show that the presence of more than two subbands leads to a qualitatively 
different MIS oscillation picture, described as a superposition of several
oscillating contributions. Under a continuous microwave irradiation, the 
magnetoresistance of triple-well systems exhibits an interference of MIS oscillations 
and microwave-induced resistance oscillations. The theory explaining these 
phenomena is presented in the general form, valid for an arbitrary number of 
subbands. A comparison of theory and experiment allows us to extract temperature 
dependence of quantum lifetime of electrons and to confirm the applicability of 
the inelastic mechanism of microwave photoresistance for the description of 
magnetotransport in multilayer systems.
\end{abstract}


\maketitle

\section{Introduction}

Studies of magnetoresistance, in particular, the investigation of Shubnikov-de Haas (SdH) 
oscillations in semiconductors and metals is an important tool for gathering information 
about band structure, quantum lifetimes of electrons and interaction mechanisms \cite{1}. 
In two-dimensional (2D) electron systems, the SdH oscillations occur because of a periodic 
modulation of electron scattering as the Landau levels consecutively pass through 
the Fermi level. With increasing temperature, when the thermal broadening of the Fermi 
distribution exceeds the cyclotron energy $\hbar \omega_c$, the SdH oscillations are strongly 
damped. In quantum wells with at least two occupied 2D subbands, the magnetoresistance 
exhibits another kind of oscillating behavior, the so-called magneto-intersubband (MIS) 
oscillations \cite{2}. These oscillations occur because of a periodic modulation of the 
probability of transitions between the Landau levels belonging to different subbands. 
The MIS oscillation peaks correspond to the subband alignment condition $\Delta = 
n \hbar \omega_c$, where $\Delta$ is the subband separation, because the isoenergetic 
(elastic) scattering of electrons between the Landau levels is maximal under this condition. 
Since the origin of the MIS oscillations is not related to the position of Landau levels with 
respect to the Fermi energy, these oscillations survive at high temperatures when 
the SdH oscillations are completely damped. Early experimental studies of MIS 
oscillations have been carried out in single quantum wells with two populated 2D subbands 
\cite{3}-\cite{5}. Recently, MIS oscillations with large amplitudes have been observed 
and investigated in two-subband systems based on double quantum wells (DQWs) \cite{6,7}. 
The DQWs appear to be the most convenient systems for experimental studies of 
this penomenon, because the two-subband occupation is attainable at relatively small 
electron densities, the electron mobility is high, and a strong tunnel coupling between 
the wells enables a high probability of intersubband scattering. The studies of MIS 
oscillations in DQWs provide information about temperature dependence of the quantum 
lifetime of electrons in the region where SdH oscillations are absent \cite{6}.

The MIS oscillations are also interesting owing to their interplay with another 
magneto-oscillatory phenomenon recently discovered in high-mobility 2D layers. If a 
2D electron system is exposed to a continuous microwave irradiation,
microwave-induced resistance oscillations (MIROs) occur, which are governed by the ratio of 
the radiation frequency $\omega$ to the cyclotron frequency $\omega_{c}$ \cite{8}. 
With increasing radiation intensity, the minima of these oscillations evolve into 
"zero-resistance states" \cite{9,10} in samples with ultrahigh electron mobility. 
Similar to MIS oscillations, MIROs originate from a periodic modulation of the 
probability of electron transitions between different Landau levels. In single-subband 
systems, such transitions occur because of electron scattering in the presence of 
microwave excitation, when electrons absorb radiation quanta and gain the energy 
necessary for the transitions. A detailed theoretical description of MIROs involves 
consideration of several microscopic mechanisms of photoresistance, which satisfactory 
describe the observed periodicity and phase of these oscillations \cite{11}-\cite{14}.
Among them, the inelastic mechanism \cite{13}, associated with a microwave-generated 
non-equilibrium oscillatory component of the isotropic part of the electron distribution 
function, dominates at low temperatures $T$, because its contribution is proportional 
to the inelastic relaxation time $\tau_{in} \propto T^{-2}$. Recently, MIROs have been 
studied in systems with two occupied subbands (DQWs) \cite{15}. It is found that 
the interplay between MIS oscillations and MIROs manifests itself as an interference 
of these kinds of oscillations, which is formally expressed as a product of the 
corresponding oscillating factors \cite{15}. The observed magnetoresistance pattern 
strongly depends on frequency $\omega$ and exhibits inversion or enhancement of certain 
groups of MIS peaks. It is established that the inelastic mechanism of microwave
photoresistance explains magnetoresistance oscillations in such two-subband systems \cite{15}.

\begin{figure}[ht]
\includegraphics[width=9cm]{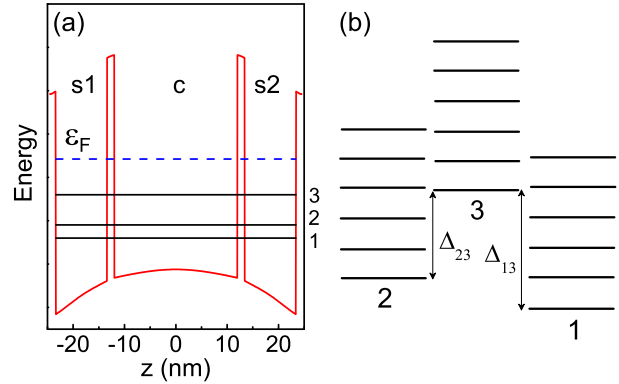}
\caption{\label{fig1} (color online) (a) Image of a triple quantum well and (b) Landau level 
        staircase for a triple quantum well with three occupied subbands (1,2,3).}
\end{figure}

Previous studies of MIS oscillations and their interference with MIROs have been 
restricted to two-subband systems. In this paper we present experimental and theoretical 
studies of these phenomena in systems with three occupied subbands formed in triple 
quantum wells (TQWs). The symmetric triple-well structure under investigation is shown 
in Fig. \ref{fig1}. The barriers dividing the wells are thin enough to have a strong tunnel
hybridization of electron states in different wells. As a result, there exist three 
subbands with different quantization energies $\varepsilon_j$ ($j=1,2,3$) and all of them 
are occupied by electrons at the chosen (high enough) electron density. We have found that 
the magnetoresistance of such systems exhibits MIS oscillations with several periods 
determined by subband separation energies $\Delta_{jj'}=|\varepsilon_j-\varepsilon_{j'}|$. 
The peculiar MIS oscillation picture is distinct from that observed in DQWs, where only 
one MIS period exists. This feature has not been mentioned in previous studies of TQWs by 
other groups \cite{16,17}, which concentrated on the regime of high magnetic fields and 
quantum Hall effect. Next, we have demonstrated that the MIS oscillation picture in 
TQWs exposed to microwave irradiation changes and depends on the radiation frequency. 
This behavior is basically similar to that in the case of DQWs described above. 
To explain the observed MIS oscillations and their interference with MIROs in 
TQWs, we generalize the magnetoresistance theory in the presence of microwave 
irradiation to the multisubband case. All experimental results are in agreement 
with theoretical caluclations involving the inelastic mechanism of photoresistance.

The paper is organized as follows. Section II presents our experimental results for 
TQWs and a theoretical description of the MIS oscillations in many-subband systems. Section 
III presents the magnetoresistance under microwave irradiation, also together 
with a theoretical description, a comparison of theory and experiment, and a discussion of the results. 
Conclusions are given in the last section. Appendix A contains details of the theoretical calculation 
of photoresistance for many-subband systems, and Appendix B describes the tight-binding 
approach for a calculation of subband spectrum and scattering rates in symmetric triple-well 
systems, with application to our samples.  

\section{MIS oscillations in triple quantum wells}

Our samples are symmetrically doped GaAs TQWs, separated by Al$_{x}$Ga$_{1-x}$As barriers, with 
a high total electron sheet density of $n_{s}~=~9 \times 10^{11}$~cm$^{-2}$ and mobilities of 
$5 \times 10^{5}$~cm$^{2}$/V s (wafer A) and $4 \times 10^{5}$ cm$^{2}$/V s (wafer B). The central well 
width is about 230 \AA\ and both side wells have equal widths of 100 \AA. The barrier thickness 
$d_b$ is 14 \AA\ (wafer A) and 20 \AA\ (wafer B). In order to make the central well populated, we 
increased its width. The estimated density in the central well is 35\%\ smaller than in the side wells.
The layers are shunted by ohmic contacts. Figure \ref{fig1} gives an image of a triple quantum well with 
three occupied subbands ($j=1,2,3$) and the staircases of Landau levels. The subband separation energies 
$\Delta_{jj'}$, which characterize the coupling strength between the quantum wells (see Appendix B), 
are presented in Table \ref{tab1}. The measurements have been carried out in a dilution refrigerator 
(low temperature MIS studies) and in a VTI cryostat using a waveguide for microwave experiments 
to deliver microwave radiation down to the sample. A conventional lock-in technique for 
magnetotransport measurements under a continuous microwave irradiation (35~GHz to 170~GHz) 
has been used. Several specimens of both van der Pauw and Hall bar geometries from both wafers 
have been studied. 

\begin{table}[ht]
\caption{\label{tab1} Subband separation energies for two samples, extracted from Fourier 
analysis of magnetoresistance.}
\begin{ruledtabular}
\begin{tabular}{lccc}
wafer & $\Delta_{12}$ (meV) & $\Delta_{23}$ (meV) & $\Delta_{13}$ (meV)\\
\hline
A & 1.4 & 3.9 & 5.3\\
B & 1.0 & 2.4 & 3.4\\
\end{tabular}
\end{ruledtabular}
\end{table}

\begin{figure}[ht]
\includegraphics[width=9cm]{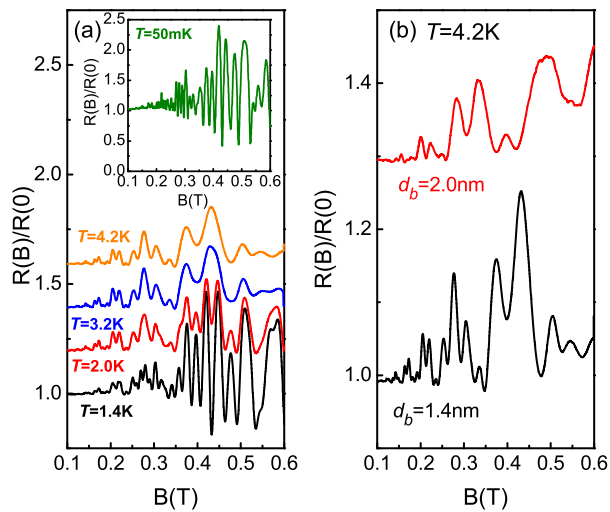}
\caption{\label{fig2}(color online) (a) Temperature dependence for MIS oscillations in a TQW with
         $d_b$~=~14~\AA\ for 1.4, 2.0, 3.2 and 4.2~K. The inset shows MIS oscillations at 50~mK,
         superimposed on low-field SdH oscillations. (b) Comparison of both wafers with $d_b$~=~20~\AA\
         (top, shifted up for clarity) and $d_b$~=~14~\AA\ (bottom) at $T~$=~4.2~K.}
\end{figure}

The description of the MIS oscillations will be focused on samples with $d_b$~=~14~\AA\ owing to
a stronger tunnel coupling which gives rise to better pronounced MIS features. In Fig. \ref{fig2} (a)
we present temperature dependence of MIS oscillations from $T~$=~1.4~K to $T~$=~4.2~K. The inset
also shows MIS oscillations at $T~$=~50~mK. For this low temperature and also at 1.4~K, the MIS oscillations
are superimposed on low-field SdH oscillations. Figure \ref{fig2} (b) demonstrates MIS oscillations for
both samples with $d_b$~=~14~\AA\ and $d_b$~=~20~\AA\ . It is obvious
that for the thicker barrier, when tunnel coupling is weaker, the subband separation $\Delta_{jj'}$
becomes smaller and the probability of intersubband transitions of electrons decreases. This is 
reflected in the periodicity and in the amplitude of MIS oscillations (see also Table \ref{tab1}).

To describe the data in more detail, we have generalized the theory of magnetoresistance in the 
systems with two occupied subbands (DQWs) \cite{6,18,19} to the case of $N$ subbands. We consider 
elastic scattering of electrons in the presence of a magnetic field under the condition of large 
filling factors (Fermi energy $\varepsilon_F$ is much larger than $\hbar \omega_c$), and apply 
the self-consistent Born approximation to describe the density of states and the linear response. 
The expression for magnetoresistance is conveniently presented in the form 
\begin{equation}
\rho_{d}= \rho^{(0)}_{d} + \rho^{(1)}_{d} + \rho^{(2)}_{d}, 
\end{equation}
where $\rho^{(0)}_{d}$ is the classical 
resistivity, $\rho^{(1)}_{d}$ is the first-order (linear in Dingle factors) quantum contribution 
describing the SdH oscillations, and $\rho^{(2)}_{d}$ is the second-order (quadratic in Dingle 
factors) quantum contribution containing the MIS oscillations. In the regime of classically 
strong magnetic fields, one obtains 
\begin{equation}
\rho^{(0)}_{d}= \frac{m}{e^2 n_s \tau_{tr}},~~~ \frac{1}{\tau_{tr}}=\frac{1}{N} \sum_j \nu^{tr}_{j},
\end{equation}
\begin{equation}
\rho^{(1)}_{d} = -{\cal T} \frac{4m}{e^2 n_s} \frac{1}{N} \sum_{j}  \nu^{tr}_{j} d_j \cos  \frac{2 \pi (\varepsilon_F-\varepsilon_j)}{\hbar \omega_c},
\end{equation}
and
\begin{equation}
\rho^{(2)}_{d}=\frac{m}{e^2 n_s} \sum_{jj'} \frac{n_j+n_{j'}}{n_s} \nu^{tr}_{jj'} 
d_j d_{j'} \cos  \frac{2 \pi \Delta_{jj'}}{\hbar \omega_c},
\end{equation}
where $m$ is the effective mass of electrons, $d_j=\exp(-\pi \nu_j/\omega_c)$ are the Dingle 
factors, ${\cal T}= X/\sinh X$ with $X=2 \pi^2 T/\hbar \omega_c$ is the thermal suppression 
factor, and $n_j$ are the partial densities in the subbands ($\sum_j n_j=n_s$). The sums are 
taken over all subbands, and since for the terms with $j=j'$ one has $\Delta_{jj'}=0$, the 
corresponding cosines in Eq. (4) are equal to 1. The subband-dependent quantum relaxation 
rates $\nu_j$ and $\nu_{jj'}$, as well as the transport scattering rates $\nu^{tr}_j$ and 
$\nu^{tr}_{jj'}$ entering Eqs. (2)-(4) are defined according to
\begin{equation}
\nu_j=\sum_{j'} \nu_{jj'},~~~\nu^{tr}_j= N \sum_{j'} \frac{n_j+n_{j'}}{2 n_s} \nu^{tr}_{jj'},
\end{equation}
and
\begin{eqnarray}
\begin{array}{c} \nu_{jj'} \\
\nu^{tr}_{jj'} \end{array}
 \left\} = \int_0^{2 \pi}
\frac{d \theta}{2 \pi} \nu_{jj'}(\theta) 
\times \right\{\begin{array}{c} 1 \\
F_{jj'}(\theta) \end{array} , \\
\nu_{jj'}(\theta)=\frac{m}{\hbar^3} w_{jj'} \left( \sqrt{ (k^2_{j} +
k^2_{j'}) F_{jj'}(\theta)} \right), \nonumber
\end{eqnarray}
where $w_{jj'}(q)$ are the Fourier transforms of the correlators of the scattering potential, 
$F_{jj'}(\theta)=1 - 2 k_j k_{j'} \cos\theta/(k_j^2 + k_{j'}^2)$, $\theta$ is the scattering angle, 
and $k_j=\sqrt{2 \pi n_j}$ is the Fermi wavenumber for subband $j$. Since $F_{jj}=1-\cos\theta$, 
the intrasubband transport rates $\nu^{tr}_{jj}$ are defined in a conventional way. The theory is 
valid if the subband separations $\Delta_{jj'}$ are large compared to the broadening energies 
$\hbar \nu_j$ and the Dingle factors are small, $d^2_j \ll 1$. In a similar way as we introduced the 
averaged transport time $\tau_{tr}$ by Eq. (2), one can introduce the averaged quantum lifetime 
$1/\tau_{q}=N^{-1} \sum_j \nu_{j}$. Application of Eqs. (1)-(6) to the particular case of three 
subbands ($N=3$, $j=1,2,3$) is straightforward.

The behavior of magnetoresistance can be illustrated within a simple model using equal 
electron densities $n_j=n_s/N$ and assuming that all $\nu^{tr}_{jj'}$ and $d_j$ are equal to 
each other, in particular, $d_j=d=\exp(-\pi /\omega_c \tau_q)$. Neglecting the SdH oscillations, 
we obtain for $N=3$
\begin{eqnarray}
\frac{\rho_d(B)}{\rho_d(0)} \simeq 1 + \frac{2}{3} d^2 \left[1+
\frac{2}{3}\cos \left(\frac{2\pi \Delta_{12}}{\hbar \omega_c} \right) \right. \nonumber \\ 
\left. + \frac{2}{3}\cos\left(\frac{2 \pi \Delta_{13}}{\hbar \omega_c} \right)+ 
\frac{2}{3}\cos\left(\frac{2\pi\Delta_{23}}{\hbar \omega_c}\right)\right].
\end{eqnarray}
The MIS oscillations are represented as a superposition of three oscillating terms determined 
by relative positions of the subband energies. Notice that this expression does not depend on 
transport rates except the one standing in the Dingle factor $d$. The approximation (7), in principle, 
can be applied for estimates to our system, since we have high total electron-sheet density and a 
strong tunnel coupling. To describe experimental magnetoresistance in detail, a more careful 
calculation based on Eqs. (1)-(6) is required.

\begin{figure}[ht]
\includegraphics[width=9cm]{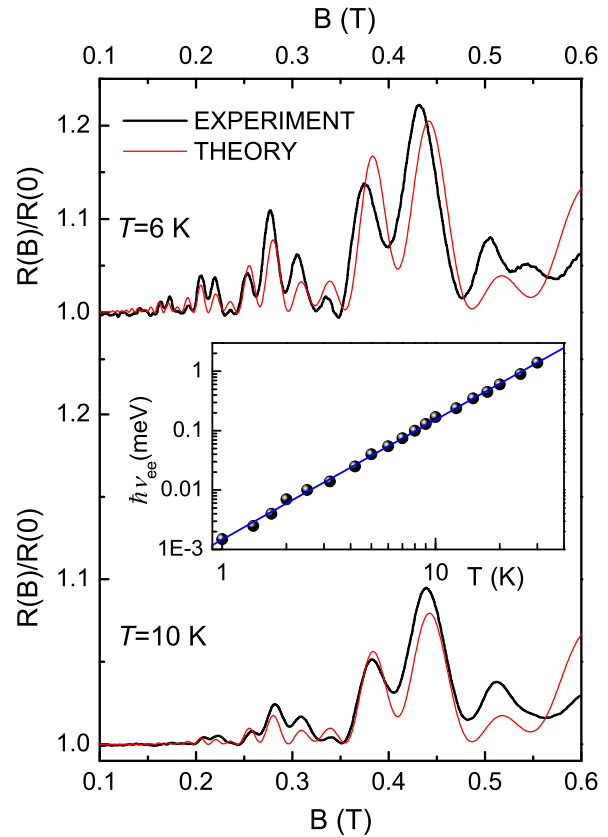}
\caption{\label{fig3}(color online) Comparison of the experimental and theoretical
        traces for a TQW with $d_b$~=~14~\AA\ at $T~$=~6~K (top) and $T~$=~10~K (bottom).
        By fitting the amplitude of MIS oscillations, electron-electron scattering rate $\nu_{ee}$ is 
        extracted (see the points in the inset). The linear fit to experimental data (line) 
        corresponds to the theoretical dependence of Eq. (8) with $\lambda = 2.2$.}
\end{figure}

We calculate the magntoresistance of our system under a simplified assumption that the 
scattering potential is essential only in the side $(s)$ wells, since the growth technology 
implies that most of the scatterers reside in the outer barriers. The correlation length 
$l_c=18.3$ nm entering the scattering potential correlator $w_{jj'}(q) \propto w_s(q)$ 
(see Appendix B for details) is determined by comparing the results of calculations to 
low-temperature magnetoresistance data for the samples with mobility $5 \times 10^{5}$~cm$^{2}$/V s. 
The averaged quantum lifetime estimated in this way is $\tau_q \simeq 3.8$ ps. The experiment shows 
a slow suppression of the MIS oscillations with temperature, which occurs owing to the contribution 
of electron-electron scattering into Landau level broadening. Though the theory presented above 
does not take this effect into account explicitly, it can be improved by replacing the quantum 
relaxation rates according to
\begin{equation}
\nu_j~ \rightarrow ~\nu_j + \nu_{ee},~~~\nu_{ee} = \lambda \frac{T^{2}}{\hbar \varepsilon_F}
\end{equation}
where $\nu_{ee}$ is the electron-electron scattering rate \cite{20,21}, the Fermi energy is 
expressed through the averaged electron density as $\varepsilon_F= \hbar^2 \pi (n_s/3)/m$, and 
$\lambda$ is a numerical constant of order unity. 
In Fig. \ref{fig3} we present a comparison of experiment and theory for two choosen temperatures, 
$T~$=~6~K and $T~$=~10~K. We have carried out this procedure for many temperatures from 
$T~$=~1~K up to 30~K, and estimated $\nu_{ee}$ by fitting the amplitudes of theoretical 
and experimental magnetoresistance traces. The effect of electron-electron scattering becomes 
essential for $T > 2$ K and strongly reduces the amplitude of the MIS oscillations at $T \sim 10$ K. 
As seen from the log-log plot in the inset to Fig. \ref{fig3}, the extracted scattering rate 
$\nu_{ee}$ follows the $T^2$ dependence, in accordance with Eq. (8). This behavior is similar 
to that observed in DQWs \cite{6}. Using $\varepsilon_{F}~\simeq~10.5$~meV, we find $\lambda = 2.2$.

\section{Influence of microwaves on MIS oscillations}

In this section, we investigate the influence of a continuous microwave irradiation on our TQW system.
We have studied power, temperature, and frequency dependence of magnetoresistance for both wafers, 
though we focus again on the samples with $d_b$~=~14~\AA\ . 

\begin{figure}[ht]
\includegraphics[width=9cm]{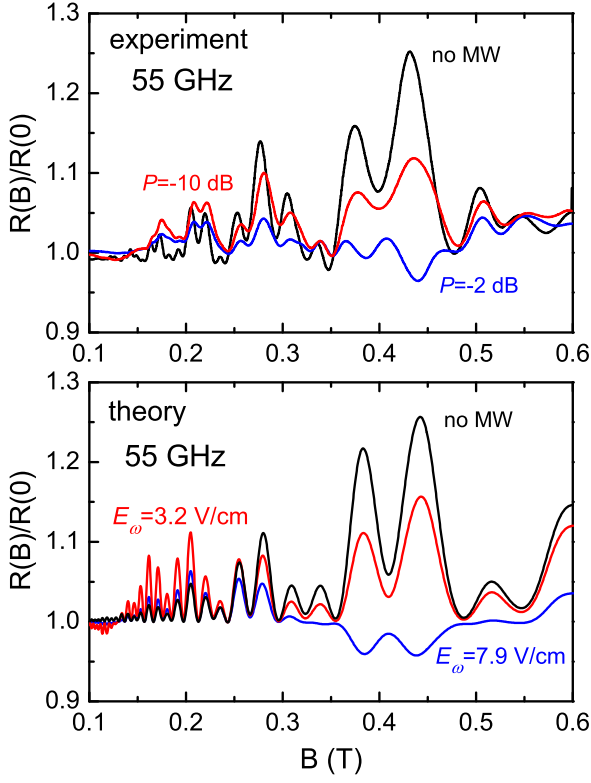}
\caption{\label{fig4}(color online) Measured (upper panel) and calculated (lower panel) magnetoresistance
        for TQW with $d_b$~=~14~\AA\ at a lattice temperature of $T~$=~4.2~K under excitation by microwaves 
        with a frequency of 55~GHz. The microwave electric field $E_{\omega}$ used for calculation of the 
        magnetoresistance (see details in the text) corresponds to a different microwave power in dB.}
\end{figure}

In the upper part of Fig. \ref{fig4} we present the magnetoresistance for different microwave powers 
at a temperature of $T~$=~4.2~K and a frequency of 55~GHz. Without microwave irradiation (no MW), 
only the MIS oscillations are visible. An increase in microwave power ($-10$~dB attenuation) 
leads to an enhancement of all MIS features for $B~<~$0.25~T and to a damping of all such 
features for $B~>~$0.35~T whereas the MIS oscillations around $B~=~$0.3~T are almost unchanged. 
A further increase in power ($-2$~dB attenuation) leads to a damping of the MIS oscillation 
amplitude for $B~<~$0.25~T, slightly increased compared to the MIS oscillation amplitude without 
microwave irradiation. The MIS features around $B~=~$0.3~T are considerably damped, while for 
0.35~T~$<~B~<~$0.5~T the MIS peaks are inverted. No polarization dependence of magnetoresistance 
has been found.

\begin{figure}[ht]
\includegraphics[width=9cm]{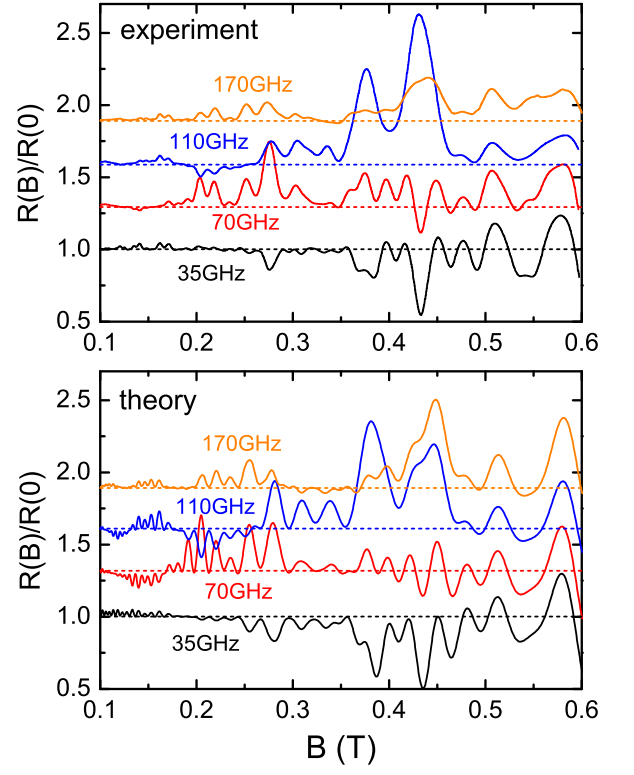}
\caption{\label{fig5}(color online) Measured (upper panel) and calculated (lower panel) magnetoresistance
        for a TQW with $d_b$~=~14~\AA\ at the lattice temperature $T~$=~1.4~K for several choosen 
        frequencies. The curves for 70, 110, and 170~GHz are shifted up for clarity.}
\end{figure}

A similar behavior, with enhanced, suppressed, or inverted MIS peaks is observed in the 
magnetoresistance measured at different microwave frequencies in the range between 35~GHz and 
170~GHz, as shown in the upper part of Fig. \ref{fig5}. A strongly modified picture of the MIS 
oscillations correlates with the microwave frequency. The features most affected by the 
microwave irradiation, strongly sensitive to its frequency, occur at $B$~=~0.27~T and 
$B$~=~0.43~T. The plots for 110~GHz and 170~GHz definitely show several regions of enhanced 
peaks and two regions of suppressed or inverted peaks (for example, the regions around 0.18 T 
and 0.34 T for 170~GHz). For 35~GHz, all the MIS oscillations above 0.2 T are 
inverted. For 35~GHz and 70~GHz, some SdH oscillations are visible in the region 
above 0.4 T. For 110~GHz and 170~GHz, when the absorption of microwave radiation in this 
region is higher, the SdH oscillations are suppressed because of the heating of the electron gas 
by microwaves.

The peculiar features of the microwave-modified magnetoresistance can be understood in terms of 
interference of the MIS oscillations with MIROs. This phenomenon is already known for two-subband 
systems \cite{15}. The theory applied for explanation of our measurements is presented below. The dissipative 
resistivity in the presence of microwaves is given by (see Appendix A for derivation of the expressions)
\begin{equation}
\rho_d^{{\scriptscriptstyle MW}}= \rho_d+\rho_{in}+\rho_{di},
\end{equation}
where the dark resistivity $\rho_d$ is described in Sec. II, while $\rho_{in}$ and $\rho_{di}$ 
are the microwave-induced contributions due to inelastic and displacement mechanisms, respectively.
At low temperatures, $\rho_{in}$ is the main contribution. It is given by
\begin{equation}
\rho_{in}=-\frac{m}{e^2 n_s} \frac{2 \tau_{tr} A_{\omega}}{N^2} \sum_{jj'} \nu^{tr}_{j} \nu^{tr}_{j'} 
d_j d_{j'} \cos \frac{2 \pi \Delta_{jj'}}{\hbar \omega_c},
\end{equation}
where
\begin{equation}
A_{\omega} = \frac{ {\cal P}_{\omega} (2 \pi
\omega/\omega_c) \sin (2 \pi \omega/\omega_c)}{1+{\cal P}_{\omega}
\sin^2(\pi \omega/\omega_c)}
\end{equation}
is an oscillating function describing MIROs, and 
\begin{equation}
{\cal P}_{\omega}=\frac{\tau_{in}}{\tau_{tr}}P_{\omega},~~P_{\omega}=\left(\frac{e E_{\omega} }{\hbar\omega}\right)^{2} 
\overline{v_F^2} (|s_+|^2+|s_-|^2)
\end{equation}
is the dimensionless function proportional to the microwave power.
In this expression, $\overline{v_F^2}=N^{-1} \sum_j v_j^2$ is the averaged Fermi velocity, 
and $\tau_{in}$ is the averaged inelastic relaxation time introduced in Appendix A. The averaged 
transport time $\tau_{tr}$ is defined in Sec. II, and the coefficients $s_{\pm}$ are given in Appendix 
A. Notice that in the case of linear polarization of radiation, and away from the cyclotron resonance, 
$|s_+|^2+|s_-|^2 \simeq (\omega^2+\omega_c^2)/(\omega^2-\omega^2_c)^2$. The contribution $\rho_{di}$ 
is presented in Appendix A. Since $\rho_{di}$ is much smaller than $\rho_{in}$, it is not taken into
account in our consideration. The influence of microwave radiation on the SdH oscillations can also be 
neglected at weak radiation power.

To get a visual description of the influence of radiation on magnetoresistance, we again use 
a simple model assuming equal partial densities $n_j=n_s/N$, equal transport scattering rates 
$\nu^{tr}_{jj'}$ and Dingle factors $d_j=d$. The magnetoresistance of three-subband system ($N=3$) 
then takes the form
\begin{eqnarray}
\frac{\rho^{\scriptscriptstyle MW}_d(B)}{\rho_d(0)} \simeq 1+\frac{2}{3}(1-A_{\omega}) d^2 \left[1+
\frac{2}{3}\cos \left(\frac{2\pi\Delta_{12}}{\hbar \omega_c}\right) \right. \nonumber \\ 
\left.+\frac{2}{3}\cos \left(\frac{2 \pi \Delta_{13}}{\hbar \omega_c}\right)+ 
\frac{2}{3}\cos\left(\frac{2\pi\Delta_{23}}{\hbar \omega_c}\right)\right], 
\end{eqnarray}
which differs from Eq. (7) only by the presence of the factor $1-A_{\omega}$. The products of 
$A_{\omega}$ by the MIS oscillation factors $\cos ( 2\pi\Delta_{jj'}/\hbar \omega_c )$ lead to 
interference oscillations of the magnetoresistance. In the regions of frequency where $A_{\omega}$ 
is negative, one expects an enhancement of the MIS peaks. If $A_{\omega}$ is positive, the peaks 
are suppressed and inverted with increasing microwave power. This is the main feature of the 
behavior we observe experimentally in Figs. \ref{fig4} and \ref{fig5}. 

A comparison of experimental results with theory based on Eqs. (10)-(12) is demonstrated in 
the lower parts of Figs. \ref{fig4} and \ref{fig5}. Apart from the known parameters used 
also in Sec. II, we apply the following estimate for the inelastic relaxation time \cite{13}: 
$\tau_{in} \simeq \hbar \varepsilon_F/T^{2}$, assuming that the relaxation is 
governed by the electron-electron interaction. The reliability of this estimate is confirmed 
in numerous experiments on magnetoresistance influenced by either microwave field \cite{15,22} 
or static electric field \cite{23,24}. To explain the experimental data, it is important to 
take into account microwave heating of the electron gas. This effect is directly visible 
in our experiment and results in a suppression of the SdH oscillations under microwave irradiation. The 
increase of the effective electron temperature over the lattice temperature also leads to 
a decrease of the inelastic relaxation time and quantum lifetimes, see Eq. (8), so the MIS 
oscillation amplitudes are expected to be suppressed as a result of electron heating. 
The electron temperature, which depends on the magnetic field, radiation frequency, and 
power, has been calculated assuming energy relaxation of electrons due to their interaction 
with acoustic phonons. 
Finally, to determine the electric field $E_{\omega}$ corresponding to our measurements, we use 
an estimate for the microwave electric field generated by our source as $10$ V/cm (at 55 GHz). 
Thus, the attenuations of $-10$ dB and $-2$ dB correspond to $E_{\omega}=3.2$ V/cm and 7.9 V/cm, 
respectively, and we applied these values for calculation of the magnetoresistance shown 
in Fig. \ref{fig4}. 

The theoretical plots in Fig. \ref{fig4} reproduce all the basic features of the 
experimental magnetoresistance traces, in particular, a suppression and inversion 
of two MIS peaks around $B=0.4$ T, because of the contribution $\rho_{in}$ with 
positive $A_{\omega}$. Notice that the non-monotonic power dependence of 
the MIS peaks around 0.2 T is explained by the interplay of MIS/MIRO interference 
and heating effects. At low radiation power the enhancement of these peaks occurs 
because of the contribution $\rho_{in}$ with negative $A_{\omega}$. At high power, 
when the saturation effect takes place \cite{13,14}, a decrease in the Dingle 
factors due to the heating-induced increase in $\nu_{ee}$ becomes more important 
and the MIS peaks are suppressed.

The expected microwave electric field in the frequency-dependent measurements shown 
in Fig. \ref{fig5} is $E_{\omega} \sim 3$ V/cm. To get a closer resemblance of 
the theoretical magnetoresistance to the experimental plots, we slightly varied 
$E_{\omega}$ around this value and obtained the best fit at $E_{\omega}=3.5$ V/cm 
for 35~GHz and 70~GHz, 4 V/cm for 110~GHz, and 2.2 V/cm for 170~GHz. The corresponding 
theoretical plots are presented in Fig. \ref{fig5}. Since the lattice temperature for these 
measurements is 1.4 K, the heating effect appears to be considerable. For 35, 70, and 110~GHz, 
the calculated electron temperature in the vicinity of the cyclotron resonance is about 
3.5~K, which is close to our experimental estimates obtained from suppression of the SdH 
oscillations. In general, a reasonably good agreement between theory and experiment at 
different frequencies suggests that the theoretical model applied for the calculations 
is reliable.

\section{Conclusions}

We have studied transport properties, including microwave photoresistance, 
of the electron systems with three occupied 2D subbands in perpendicular 
magnetic fields. Such systems are realized in TQWs with high enough electron 
density. As we have demonstrated, both experimentally and theoretically, 
the magnetoresistance of TQWs is qualitatively different from that for 
single-subband and two-subband systems, and contains a superposition of 
three oscillating terms whose frequencies are given by the subband separation 
energies $\Delta_{12}$, $\Delta_{13}$, and $\Delta_{23}$. This occurs because 
the quantum contribution to the resistivity is essentially determined by electron 
scattering between the Landau levels of different subbands. Therefore, there 
exist MIS oscillations of resistivity, and the picture of these oscillations 
becomes complicated in the systems with more than two occupied subbands. We have 
presented a theoretical description of such oscillations by generalizing the 
theory of quantum magnetoresistance to the multisubband case, and obtained a good 
agreement with the experiment. 

Similar as in single-subband and two-subband systems, the quantum contribution 
to the resistivity decreases with increasing temperature $T$ because of the decrease 
in quantum lifetime due to enhanced contribution of electron-electron scattering. By 
measuring the amplitude of the MIS oscillations at different temperatures up to 30 K, 
we have established that the temperature dependence of electron-electron scattering rate 
follows the theoretically predicted $T^2$ law, see Eq. (8). The numerical
constant $\lambda$ in this dependence, $\lambda=2.2$, is close enough to those 
determined from the MIS oscillations in two-subband systems, both in double quantum 
wells \cite{6} ($\lambda=3.5$) and in single quantum wells \cite{25} ($\lambda=2.6$).
Therefore, one may conclude that the influence of electron-electron scattering on the 
quantum lifetime of electrons is not very sensitive to the number of occupied subbands. 

The TQWs exposed to a continuous microwave irradiation demonstrate dramatic changes 
in magnetoresistance. The effect of microwaves is understood as a result of interference 
of MIS oscillations and microwave-induced resistance oscillations. A similar effect 
takes place for two-subband systems in double quantum wells \cite{15}, where it is 
easier recognizable owing to a simpler picture of the MIS oscillations. To describe 
our observations, we have developed a theory of magnetoresistance of multisubband 
systems under microwave irradiation, and applied it to our three-subband systems. 
Among the mechanisms of microwave photoresistance, the inelastic mechanism is found
to be responsible for the observed magnetoresistance features. In spite of several 
approximations of the theory, in particular, those for description of elastic scattering 
(see Appendix B), we have obtained a good agreement between theoretical and experimental 
magnetoresistance traces by using an established estimate for the inelastic relaxation time. 

In summary, our investigation of low-field magnetotransport in three-subband systems 
both with and without microwave excitation is a useful step towards understanding 
the influence of energy spectrum and scattering mechanisms on the transport properties 
of low-dimensional electrons.

{\it Acknowledgements:} The authors thank to M. A. Zudov, I. A. Dmitriev, S. Vitkalov, and S. A. 
Studenikin for helpful discussions. This work was supported by COFECUB-USP (project number 
U$_{c}$~109/08), CNPq, FAPESP and with microwave facilities from ANR MICONANO.

\appendix

\section{Microwave photoresistance of a many-subband system}

In the presence of electromagnetic radiation (microwaves) of frequency $\omega$ and under a dc 
excitation, one can derive the quantum Boltzmann equation for electrons in a magnetic field by 
using a transition to the moving coordinate frame, in a similar way as for the single-subband 
system (see \cite{14} and references therein). This leads to the kinetic equation for the Wigner 
distribution function $f_{j \varepsilon \varphi}$, which depends on the subband index $j$, energy 
$\varepsilon$, and angle $\varphi$ of the electron momentum:
\begin{eqnarray}
\omega_c \frac{\partial f_{j \varepsilon \varphi}}{ \partial \varphi}  = \sum_{j'} \int_0^{2 \pi} \frac{d \varphi'}{2 \pi} \nu_{jj'}(\varphi-\varphi') \sum_{n} [J_n (\beta_{jj'})]^2 \nonumber \\ 
\times D_{j'}(\varepsilon+n \omega +\gamma_{jj'}) [f_{j' ,\varepsilon+n \omega +\gamma_{jj'}, \varphi'} - f_{j \varepsilon \varphi}]+ J_{in}. 
\end{eqnarray}
Notice that in this Appendix we use the system of units where $\hbar=1$. In the kinetic equation, we 
introduced the dimensionless (normalized to its zero field value) density of states $D_{j}(\varepsilon)$. 
Next, $J_n(x)$ is the Bessel function, $J_{in}$ is the collision integral describing inelastic scattering, 
and $\nu_{jj'}$ are the scattering rates defined in Sec. II.
The other quantities standing in Eq. (A1) are
\begin{eqnarray}
\beta_{jj'}(\varphi,\varphi')=\frac{eE_{\omega}}{\sqrt{2}\omega} \left|s_- (v_j e^{i\varphi}-v_{j'} 
e^{i\varphi'})\right. \nonumber \\
+\left. s_+ (v_j e^{-i\varphi}-v_{j'} e^{-i\varphi'}) \right|, 
\end{eqnarray}
and
\begin{eqnarray}
\gamma_{jj'}(\varphi,\varphi')=\frac{e}{2i \omega_c} [ E_{-} (v_j e^{i\varphi}-v_{j'} e^{i\varphi'}) \nonumber \\
- E_{+} (v_j e^{-i\varphi}-v_{j'} e^{-i\varphi'})], 
\end{eqnarray}
where $E_{\pm}=E_{x} \pm i E_{y}$, ${\bf E}=(E_x,E_y)$ is the dc field strength, $E_{\omega}$ is the 
strength of microwave electric field (related to the incident microwave field strength $E_i$ in 
vacuum as $E_{\omega} =E_i/\sqrt{\epsilon^*}$, see below), and $v_j=p_j/m$ are the subband-dependent 
Fermi velocities. The factors $s_{\pm}$ describe polarization of the radiation and account for 
electrodynamic effects \cite{26,27}. For the case of linear polarization, 
\begin{equation}
s_{\pm}=\frac{1}{\sqrt{2}} \frac{1}{\omega \pm \omega_c + i \omega_p},
\end{equation}   
where $\omega_p=2 \pi e^2 n_s/ m c \sqrt{\epsilon^*}$ is the plasma frequency, 
$\sqrt{\epsilon^*}=(\sqrt{ \epsilon_{vac} }+ \sqrt{\epsilon_d})/2$, $\epsilon_{vac}=1$ is the 
dielectric permittivity of vacuum, and $\epsilon_d$ is the dielectric permittivity of the medium 
surrounding the quantum wells.

The distribution function can be expanded in the angular harmonics: 
$f_{j \varepsilon \varphi}=\sum_l f_{j \varepsilon l} e^{i l \varphi}$. 
The density of dissipative electric current in the 2D plane, ${\bf j}=(j_x,j_y)$, 
is determined by the $l=1$ harmonic: 
\begin{equation}
j_- \equiv j_x - i j_y =\frac{e}{\pi} \sum_j p_j \int d \varepsilon D_{j}(\varepsilon) f_{j \varepsilon 1},
\end{equation}
In the regime of classically strong magnetic fields, the anisotropic part of the 
distribution function is expressed through the isotropic (angular-independent) 
part $f_{j \varepsilon} \equiv f_{j \varepsilon l}$ for $l=0$. This leads to the 
expression for the current in the form
\begin{eqnarray}
j_- = \frac{e}{i \pi \omega_c} \sum_{jj'} p_j \int d \varepsilon D_{j}(\varepsilon) 
\int_0^{2 \pi} \frac{d \varphi}{2 \pi} e^{-i \varphi} \int_0^{2 \pi} 
\frac{d \varphi'}{2 \pi} \nonumber \\
\times \nu_{jj'}(\varphi-\varphi') \sum_{n} [J_n(\beta_{jj'})]^2  D_{j'}(\varepsilon+n \omega +\gamma_{jj'}) \nonumber \\
\times [f_{j' \varepsilon+n \omega +\gamma_{jj'}} - f_{j \varepsilon} ].
\end{eqnarray}
The response to $E_{-}$ [$j_-= \sigma_d E_{-}$] gives the symmetric part of 
dissipative conductivity considered below. The resistivity is then given by 
$\rho^{\scriptscriptstyle MW}_d=\sigma_d/\sigma^2_{\bot}$, where $\sigma_{\bot}=
e^2 n_s/m \omega_c$ is the classical Hall conductivity. 

The isotropic part of the distribution function can be represented in the form 
\begin{equation}
f_{j \varepsilon}= f^{(0)}_{\varepsilon} - i \omega \frac{\partial 
f^{(0)}_{\varepsilon}}{\partial \varepsilon} g_{j \varepsilon}.
\end{equation} 
where $f^{(0)}_{\varepsilon}$ is a slowly varying function of energy, which is close 
to a quasi-equilibrium (heated Fermi) distribution, while $g_{j \varepsilon}$ is a 
rapidly oscillating (periodic in $\hbar \omega_c$) function, which is also represented 
as $g_{j \varepsilon}=\sum_k g_{j k} \exp(2 \pi i k \varepsilon/\omega_c)$. 
After a substitution of expression (A7) into Eq. (A6), the contribution coming from 
$f^{(0)}_{\varepsilon}$ produces the "dark" resistivity and its modification 
by the microwaves due to displacement mechanism. The microwave modification 
of the resistivity due to the inelastic mechanism originates from the term proportional 
to $g_{j \varepsilon}$. In the following, we search for the response linear in 
the dc field and quadratic in the microwave field. This is done by expanding the Bessel 
functions in powers of $\beta_{jj'}$ and retaining only the lowest-order terms. 
Also, we use the lowest-order expansion of the density of states in the Dingle factors, 
$D_{j}(\varepsilon) \simeq 1- 2 d_j \cos [2 \pi(\varepsilon-\varepsilon_j)/\omega_c]$, 
so only the lowest oscillatory harmonics ($k=\pm 1$) are relevant.
The angular and energy averaging in Eq. (A6) in this case are carried out analytically, 
and one gets the expression for the dark resistivity $\rho_d$ (Sec. II), as well 
as the microwave-induced contributions $\rho_{di}$ and $\rho_{in}$:
\begin{eqnarray}   
\rho_{di}=-\frac{m}{e^2 n_s} P_{\omega} 
\left[ \sin^2 \frac{\pi \omega}{\omega_c} + \frac{\pi \omega}{\omega_c} 
\sin \frac{2\pi \omega}{\omega_c} \right] \nonumber \\ 
\times \frac{N}{2} \sum_{jj'} \left(\frac{n_j+n_{j'}}{n_s}\right)^2 
\nu^*_{jj'} d_j d_{j'} \cos \frac{2 \pi \Delta_{jj'}}{\omega_c},
\end{eqnarray}  
where $P_{\omega}$ is defined by Eq. (12),
\begin{equation}   
\nu^*_{jj'} = \int_0^{2 \pi} \frac{d \theta}{2 \pi} \nu_{jj'}(\theta) [F_{jj'}(\theta)]^2,
\end{equation}  
and
\begin{eqnarray}     
\rho_{in}=-\frac{m}{e^2 n_s} \frac{2 \pi \omega}{\omega_c} \sum_{jj'} \frac{n_j+n_{j'}}{n_s} \nu^{tr}_{jj'} \nonumber \\
\times \left[ g_{j'1}d_j \exp \left(\frac{2 \pi i \varepsilon_j}{\omega_c} \right) + c.c. \right].
\end{eqnarray}  
One can find $g_{j \varepsilon}$ from the isotropic part of Eq. (A1) by using the relaxation 
time approximation for the isotropic part of the inelastic collision integral: 
\begin{equation}
J_{in}= -\frac{f_{j \varepsilon}-f^{(0)}_{\varepsilon}}{\tau_j^{in}}.
\end{equation} 
This approximation is valid for small deviations $f_{j \varepsilon}-f^{(0)}_{\varepsilon}$, 
and is justified in a similar way as for the single-subband systems, based on a
linearization of the collision integral for electron-electron scattering \cite{13}.
After substitution of expression (A11) into Eq. (A1), one can obtain a system of linear 
equations for $g_{j k}$, which is easily solved under a reasonable condition that 
the intersubband scattering dominates over the inelastic one: $\nu_{jj'} \gg 1/\tau_j^{in}$ 
($j \neq j'$). This gives subband-independent harmonics $g_{j k}=g_k$, in particular,
\begin{eqnarray}
g_1= \frac{P_{\omega} \tau_{in} \sin( 2 \pi \omega/\omega_c)}  
{1+ P_{\omega} (\tau_{in}/\tau_{tr}) \sin^2(\pi \omega/\omega_c)} \nonumber \\
\times \frac{1}{2 N} \sum_j \nu^{tr}_j 
d_j \exp \left(- \frac{2\pi i \varepsilon_j}{\omega_c} \right),
\end{eqnarray} 
where the averaged inelastic relaxation time is defined according to 
$1/\tau_{in} = N^{-1} \sum_j 1/\tau^{in}_{j}$. The denominator in Eq. (A12) describes the 
effect of saturation. A substitution of the result Eq. (A12) into Eq. (A10) leads to Eq. (10).

\section{Electron spectrum and scattering rates in a triple-well system}

To describe the scattering rates, we employ the wave functions and electron energies 
in the subbands found from the tight-binding Hamiltonian \cite{28} using the expansion 
of the wave function $\psi(z)= \sum_{i} \varphi_i F_i(z)$ in the basis of single-well 
orbitals $F_i(z)$ ($i=1,2,3$ numbers the left, central, and right well, respectively). 
This leads to the matrix equation for the coefficients $\varphi_i$: 
\begin{equation}
\left(\begin{array}{ccc}
\varepsilon^{(0)}_1 - \varepsilon & -t_{12} & 0 \\
-t_{12} & \varepsilon^{(0)}_2- \varepsilon & -t_{23}\\
0 & -t_{23} & \varepsilon^{(0)}_3- \varepsilon \end{array} \right)
\left( \begin{array}{c} \varphi_1 \\ \varphi_2 \\ \varphi_3 \end{array} \right) =0,
\end{equation}
where $\varepsilon^{(0)}_i$ are the single-well quantization energies and $t_{ii'}$ are the 
tunneling amplitudes. For the case of symmetric TQWs ($\varepsilon^{(0)}_1=\varepsilon^{(0)}_3 
\equiv \varepsilon_s$, $\varepsilon^{(0)}_2 \equiv \varepsilon_c$, $t_{12}=t_{23} \equiv t$) 
the energies of the three subbands, $\varepsilon_j$, are \cite{28}
\begin{eqnarray}
\varepsilon_1=(\varepsilon_c+\varepsilon_s)/2 - \Lambda, \nonumber \\ 
\varepsilon_2=\varepsilon_s, \\
\varepsilon_3=(\varepsilon_c+\varepsilon_s)/2 + \Lambda, \nonumber \\
\Lambda=\sqrt{(\varepsilon_c-\varepsilon_s)^2/4 +2t^2}. \nonumber
\end{eqnarray}
The corresponding eigenstates are expressed through the single-well orbitals as 
$\psi_j(z)=\sum_{i} \chi_{ij} F_i(z)$. The matrix $\chi_{ij}$ is given by
\begin{equation}
\chi_{ij}= \left(\begin{array}{ccc}
C_1 t/(\varepsilon_s-\varepsilon_1) & 1/\sqrt{2} &  C_3 t/(\varepsilon_s-\varepsilon_3) \\
C_1 & 0 & C_3\\
C_1 t/(\varepsilon_s-\varepsilon_1) & - 1/\sqrt{2} & C_3 t/(\varepsilon_s-\varepsilon_3) \end{array} \right),
\end{equation}
where $C_{1,3}=\left[1+2t^2/(\varepsilon_s-\varepsilon_{1,3})^2 \right]^{-1/2}$. This matrix is 
composed from the three columns of $\varphi_i$ for the states $j=1,2,3$. 

The parameters of the tight-binding model can be extracted from the subband gaps found experimentally. 
By setting $\varepsilon_s$ as the reference energy, one has:
\begin{equation}
\varepsilon_c=\Delta_{23}-\Delta_{12},~~t=\sqrt{\Delta_{23} \Delta_{12}/2}.
\end{equation}
For our samples (wafer A) we obtain $\varepsilon_c=2.5$ meV and $2 t=3.35$ meV. Using the total density 
$n_s=9 \times 10^{11}$ cm$^{-2}$, one can find the subband densities $n_1=3.62 \times 10^{11}$ cm$^{-2}$, 
$n_2=3.23 \times 10^{11}$ cm$^{-2}$, and $n_3=2.14 \times 10^{11}$ cm$^{-2}$. The electron density in 
each side well is $n_{side}=\sum_j \chi^2_{1j} n_j=\sum_j \chi^2_{3j} n_j=3.23 \times 10^{11}$ cm$^{-2}$, 
and the electron density in the central well is $n_{cent}=\sum_j \chi^2_{2j} n_j= 2.53 \times 10^{11}$ cm$^{-2}$. 
 
The random scattering potential acting on electrons is $V({\bf r},z)$, where ${\bf r}$ is the 
in-plane coordinate vector, and $z$ is the coordinate in the growth direction. The matrix elements 
of this potential, $V_{jj'}({\bf r})$, are expressed through the effective 2D potentials in the 
layers, introduced as $V_i({\bf r})= \int dz |F_i(z)|^2 V({\bf r},z)$. Accordingly, the correlators 
of the potentials are written through the correlators of $V_i({\bf r})$:
\begin{eqnarray}
W_{jj'}(|{\bf r}-{\bf r'}|) \equiv \left< \left< V_{jj'}({\bf r}) V_{j'j}({\bf r}') \right> \right> \nonumber \\
= \sum_{i i'} \chi_{ij} \chi_{ij'} \chi_{i'j'} \chi_{i'j} \widetilde{W}_{ii'}( |{\bf r}-{\bf r'}|).
\end{eqnarray}
The product of the factors $\chi$ determines the overlap of the electron wave functions, 
while the factors $\widetilde{W}_{ii'} (|{\bf r}-{\bf r'}|) \equiv 
\left< \left< V_i({\bf r}) V_{i'}({\bf r}') \right> \right>$ describe intralayer 
$i=i'$ and interlayer $i \neq i'$ potential correlations. In TQWs one can neglect 
interlayer correlations between the potentials of the side wells because of  
a large distance between these wells: $\widetilde{W}_{13} \simeq 0$. Then, symmetric 
TQWs are characterized by three correlators: $W_s=\widetilde{W}_{11}=\widetilde{W}_{33}$, 
$W_c=\widetilde{W}_{22}$, and $W_{sc}=\widetilde{W}_{12}=\widetilde{W}_{23}=
\widetilde{W}_{21}=\widetilde{W}_{32}$. A reasonable approximation for our TQWs is 
to assume that the effective potential in the central well, $V_2({\bf r})$, is much weaker 
than the side-well potentials. In this case, $W_{c}$ and $W_{sc}$ can be neglected compared 
to $W_s$, and Eq. (B5) is rewritten as
\begin{equation}
W_{jj'}(|{\bf r}-{\bf r'}|) \simeq 2 \chi^2_{1j} \chi^2_{1j'} W_s (|{\bf r}-{\bf r'}|).
\end{equation}
The spatial Fourier transform $w_{jj'}(q) = \int d {\bf r} \exp \left( -i {\bf q} \cdot {\bf r} \right) 
W_{jj'}(|{\bf r}|)$ determines all scattering rates according to Eqs. (5) and (6). In the approximation 
Eq. (B6), $w_{jj'}(q) \simeq  \chi^2_{1j} \chi^2_{1j'} w_{s}(q)$, where $w_{s}(q)$ 
is the spatial Fourier transform of $W_s$. The concrete form of the function $w_s(q)$ depends 
on the nature of the scatterers, their distribution in the structure, and on the TQW potential 
which determines the shape of $F_i(z)$. In the case of long-range scattering potential, the 
magnetoresistance $\rho_d(B)/\rho_d(0)$ is weakly sensitive to this form, but essentially depends 
on the effective correlation length $l_c$ which defines the scale of the $q$-dependence. 
In our calculations, we use a model $w_s(q) \propto \exp(-l_c q)$.

\end{document}